\documentclass[twocolumn,english,prl]{revtex4-1}
\usepackage{amsmath}
\usepackage{graphicx}
\usepackage{hyperref}
\usepackage{color}

\newcommand{\red}[1]{{#1}}

\begin{document}

\title{Topological charge pumping in a one-dimensional optical lattice}
\author{Lei Wang$^{1,2}$, Matthias Troyer$^{1}$ and Xi Dai$^{2}$}

\affiliation{$^{1}$Theoretische Physik, ETH Zurich, 8093 Zurich, Switzerland}

\affiliation{$^{2}$Beijing National Lab for Condensed Matter Physics and Institute of Physics, Chinese Academy of Sciences, Beijing 100190, China}


\begin{abstract}
A topological charge pump \cite{Thouless:1983p23000} transfers charge in a quantized fashion. The quantization is stable against the detailed form of the pumping protocols and external noises and shares the same topological origin as the quantum Hall effect. We propose an experiment setup to realize topological charge pumping of cold fermionic atoms in a one-dimensional optical lattice. The quantization of the pumped charge is confirmed by first-principle simulations of the dynamics of uniform and trapped systems. Quantum effects are shown to be crucial for the topological protection of the charge quantization. Finite-temperature and non-adiabatic effect on the experimental observables are discussed. Realization of such a topological charge pump serves as a firm step towards exploring topological states and non-equilibrium dynamics using cold atoms.
\end{abstract}
\maketitle

\paragraph{Introduction} Charge pumping is a standard method to generate steady current in solid-state circuits \cite{Shilton:1996p66663,Switkes:1999p64768, Blumenthal:2007p64769, Kaestner:2008p64776} through adiabatically and periodically time-varying potentials.  The effect bears a similarity to the famous Archimedes' screw \cite{Altshuler:1999p64775}, where water is pumped by a rotating spiral tube. However, quantum physics offers a more intriguing phenomenon: the \emph{quantum} charge pumping, where the charge transferred in each pumping cycle is exactly quantized.   
Thouless \cite{Thouless:1983p23000} has shown that the one-dimensional (1D) quantum charge pump shares the same topological origin as the two-dimensional (2D) quantum Hall effect (QHE) \cite{Klitzing:1980p64777}. The amount of pumped charge can be expressed by the Chern number of a 2D QHE Hamiltonian \cite{Thouless:1982p24208}. In other areas of condensed matter physics the theory of quantized charge pumping also lays a firm foundation for the modern theory of polarization of crystalline solids \cite{KingSmith:1993p61262, Resta:1994p61451}, the theory for $Z_{2}$ spin pump \cite{Shindou:2005df,Fu:2006p3991}, 
and inspired the theoretical connection \cite{Qi:2008p12545} between the 3D $Z_{2}$ topological insulators and the 4D quantum Hall effect \cite{Zhang:2001p35316}. 
The word \emph{quantum} in the quantum charge pumping has two-fold meanings. First the pumped charge is quantized. Second, one actually relies on the quantum mechanics (thus the concept of Berry phase and energy gap) for the topological protection of the quantized charge.

Clean and highly tunable cold atoms system provides an opportunity to realize and detect this topological charge pumping effect. Specifically, advances in constructing optical superlattice structure \cite{Peil:2003p61264, Folling:2007p55101, Trotzky:2008p464} and nonequilibrium control of lattice intensity and phases \cite{Atala:2012p70234} allow the realization of a charge pumping setup, which we will propose in this Letter. In situ detection with the single-site resolution \cite{Bakr:2010p33127, Sherson:2010p21292, Weitenberg:2011p29386} allows the detection of topological charge pumping. The equivalence of 1D topological charge pumping and the 2D quantum Hall effect connects our proposal to recent efforts of exploring topological quantum phases with synthetic gauge field \cite{Lin:2009p15747, Aidelsburger:2011p67955, Struck:2012p56790, JimenezGarcia:2012p69271}and spin-orbit couplings \cite{Lin:2011p56010, Wang:2012p61874, Cheuk:2012p61844}, where one of the landmarks is to realize the quantum Hall effect \cite{Klitzing:1980p64777} and topological insulator \cite{Hasan:2010p23520} state in atomic quantum gases. 

In this Letter, we consider the topological charge pumping of cold fermions in a 1D optical lattice potential. First we show the proposed potential indeed realizes the topological charge pumping by calculating its Chern number and \emph{ab initio} simulation of the pumping process. Compared with the corresponding classical dynamics, we show that  quantum effects are crucial for the topological protection of the quantized charge pumping. We then consider the effect of a harmonic trap and predict the topological quantization of the center of mass of the cloud in realistic experimental situations.

\begin{figure}[tbp]
\centering
    \includegraphics[width=8.5cm]{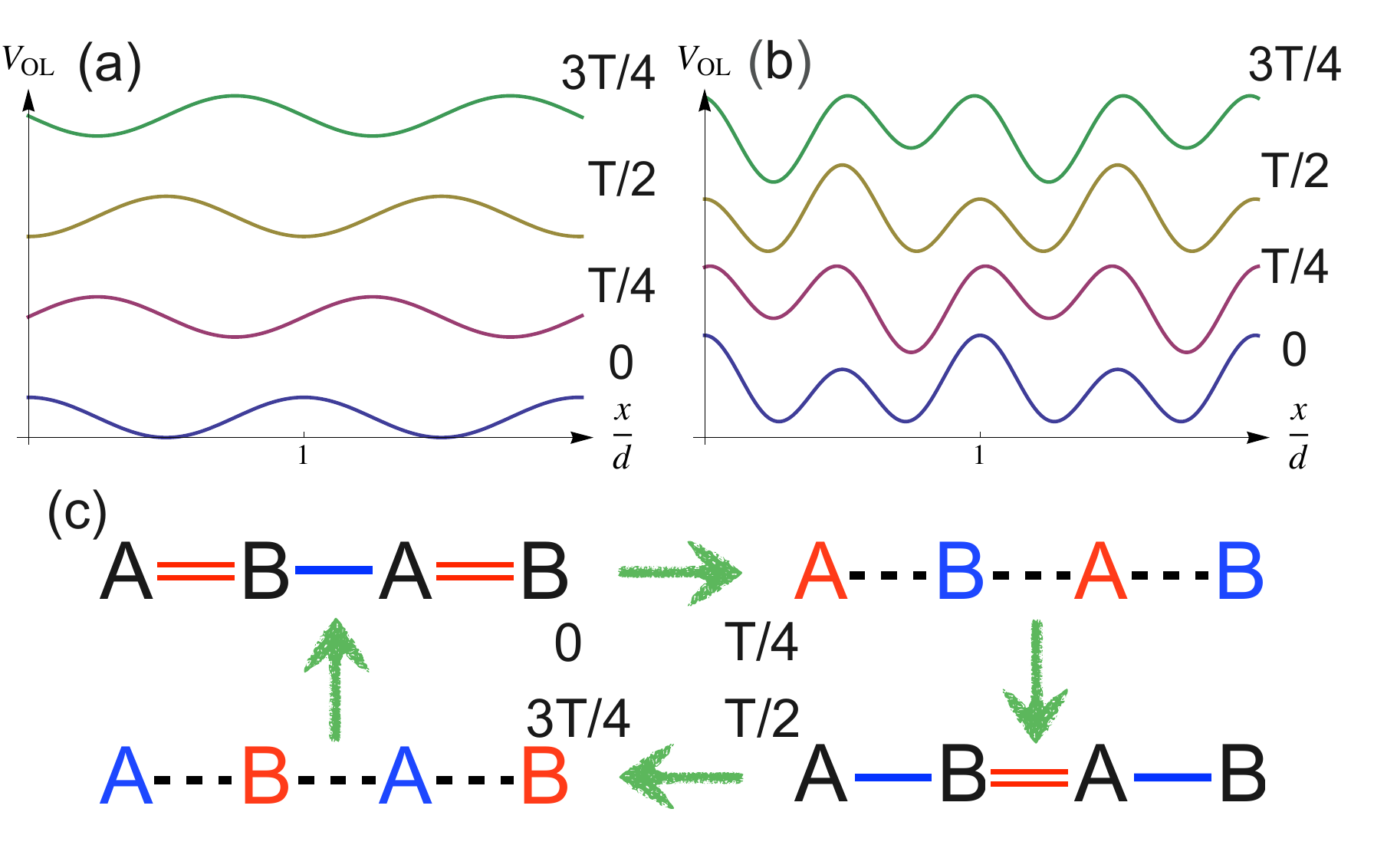}
\caption{Two topological equivalent pumping lattices. (a). A sliding lattice ($V_{1}=0,V_{2}=1E_{R}$) (b). A continuous Rice-Mele pump \cite{Rice:1982p61164} ($V_{1}=2E_{R},V_{2}=1E_{R}$), where the dimerization of hopping amplitudes and onsite energies are modulated periodically. At $t=0$ and $t=T/2$ different topological phases of the Su-Schrieffer-Heeger lattice \cite{Su:1979p972} are realized. (c). A tight-binding schematic view the pumping process \red{in} (b). The two pumping processes (a) and (b) are topologically equivalent for the lowest band and can be adiabatically connected, see the main text. }
\label{fig:lattice}
\end{figure}

Our proposal is based on a time-dependent 1D optical superlattice of the form \cite{Peil:2003p61264, Folling:2007p55101, Trotzky:2008p464, Atala:2012p70234}:
\begin{eqnarray}
V_\mathrm{OL}(x,t) =  V_{1} \cos^2\left(\frac{2 \pi x}{d} \right) +V_{2} \cos^2\left( \frac{\pi x}{d} - \red{\varphi}(t) \right).
\label{eq:lattice}
\end{eqnarray}
We use the lattice constant $d$ as the unit of length and the recoil energy $E_{R}=\frac{\hbar^{2}\pi^{2}}{2md^{2}}$ as the   unit of energy, where $m$ is the mass  of the atom. Such superlattices have been experimentally realized in Refs. \cite{Peil:2003p61264, Folling:2007p55101,Trotzky:2008p464, Atala:2012p70234}. \red{The lattice strengths $V_{1}$ and $V_{2}$ and the phase factor $\varphi$ can be tuned dynamically}. 


We propose to vary the relative phase linearly with time 
\begin{equation}
\varphi (t)=\frac{\pi t}{T}
\end{equation}
The lattice then changes in time with a period $T$. In the absence of the static short wave length lattice controlled by the $V_{1}$ term, the $V_2$ term describes a sliding lattice shown in Fig.~\ref{fig:lattice}(a). Including the $V_{1}$ term, one realizes the Rice-Mele model \cite{Rice:1982p61164} in a continuous space setup, as illustrated in Fig.~\ref{fig:lattice}(b-c). This can easily be seen by expanding the $V_{2}$ term (neglecting spatial independent constants) into two oscillating terms individually controlling the dimerized hopping amplitudes and  the sub-lattices energy offsets:
\begin{eqnarray}
V_{2}\cos\left(\frac{2\pi t}{T}\right) \cos^{2}\left(\frac{\pi x}{d}\right)   + V_{2}\sin\left(\frac{2\pi t}{T}\right)\cos^{2}\left(\frac{\pi x}{d}- \frac{\pi}{4}\right) 
\label{eq:expansion}
\end{eqnarray}
At time $t=0$ and $t=T/2$ only the first term is non-zero, realizing the  Su-Schrieffer-Heeger (SSH) model \cite{Su:1979p972}. This model exhibits two topologically distinguishable phases which are protected by inversion symmetry. The second term of Eq.~(\ref{eq:expansion}) breaks this inversion symmetry and smoothly connects the two phases of the SSH model.  At $t=T/4$ the system has uniform hopping amplitude but different  onsite energies at two sublattices. At $t=T/2$, the system enters a different topological phase of the SSH model than at $t=0$. Since the gap of the Hamiltonian does not close during the  pumping process, we can define a topological index associated with the pumping process. This index is just the Chern number of a 2D QHE Hamiltonian and gives the charge pumped during one cycle. \red{The continuum potential Eq. (\ref{eq:lattice}) interpolates between the sliding potential and the Rice-Mele model. 
They are topologically equivalent since one could adiabatically switch on the $V_{1}$ term without closing the gap }\footnote{This equivalence is only limited to the lowest band since the band gap between higher bands  closes. }.

\begin{figure}[tbp]
\centering
    \includegraphics[width=8.5cm]{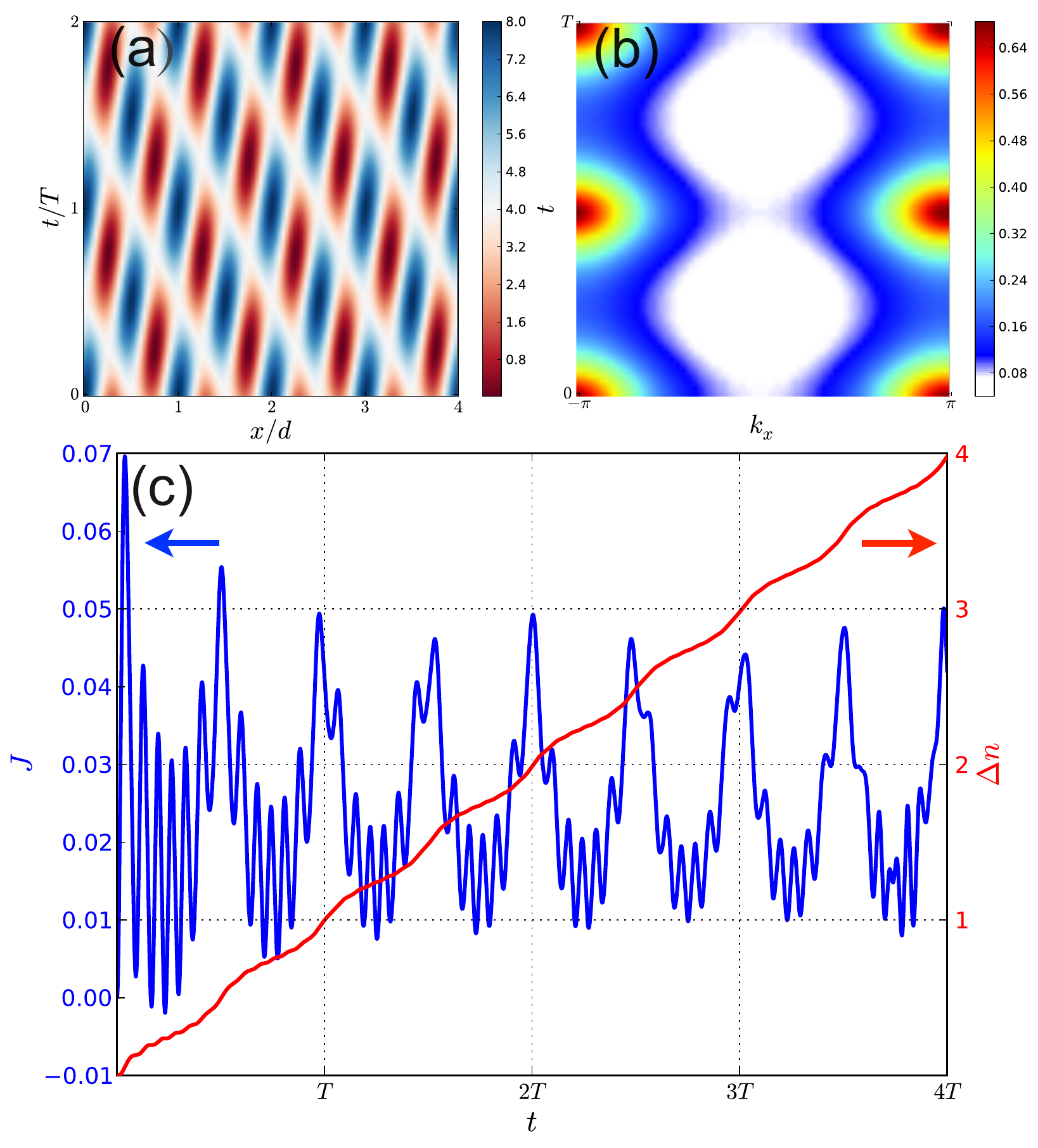}
\caption{(a). Spatial-temporal structure of the optical lattice Eq. (\ref{eq:lattice}) for $V_{1}=V_{2}=4E_{R}$. (b). Berry curvature distribution of the lowest band of $H(k_{x},t)$. Integration over the Brillouin zone shows the Chern number equals to \red{$1$}. (c). Pumped charge and total current Eq.~(\ref{eq:pumpedcharge}) of an infinite sized system with $T=40\hbar/E_{R}$ and one particle per unit-cell. The pumped charge is quantized at full pumping cycles. }
\label{fig:berry}
\end{figure}

\paragraph{Infinite system }
We first consider topological charge pumping of spinless fermions in an infinite  periodic system with Hamiltonian
\begin{equation}
H(x,t) = -\frac{\hbar^{2}}{2m}\nabla^{2} + V_\mathrm{OL}(x,t) \label{eqn:ham}.
\end{equation}
There are several advantages to working with a continuum model instead of a tight-binding (TB) lattice model like in Ref. \cite{Qian:2011p41321}. Continuum models apply to a broader range of experimental situations including shallow optical lattices. Multi-band effects are fully included in our calculations~\cite{Modugno:2012p61807}. 
A continuum model also allows us to directly compare with  classical pumping dynamics, and to demonstrate the importance of quantum effects for the topological protection of the pumped charge.

Figure~\ref{fig:berry}(a) shows the spatial-temporal structure of $V_\mathrm{OL}(x,t)$ for $V_{1}=V_{2}=4E_{R}$, which we will refer to as the Rice-Mele pumping potential in following. Performing a Fourier transform of $H(x,t) $ we obtain the Bloch Hamiltonian $H(k_{x},t)$, which satisfies the periodicity conditions $H(k_{x}+2\pi,t)=H(k_{x},t)$ and $H(k_{x},t+T) = H(k_{x},t)$. Since there is always a gap to higher bands we can calculate the Chern number of the lowest band of $H(k_{x},t)$ as if it was a two-dimensional Hamiltonian \cite{Xiao:2010p19709, Fukui:2005p38594}.  Fig.~\ref{fig:berry}(b) shows the Berry curvature distribution in the $k_{x}-t$ space. Integration of the Berry curvature over the Brillouin zone shows that the Chern number is equal to \red{$1$}. 

Thouless showed \cite{Thouless:1983p23000} that at zero temperature under the adiabatic approximation the pumped charge equals the Chern number for a filled band.  We now proceed to simulate the pumping process and directly calculate the pumped charge. For an infinite system the pumped charge is defined through the integration of the total current (See supplemental materials for details),  
\begin{equation}
\Delta n (t) = \int_{0}^{t} \!\mathrm{d} t^{\prime}\, J(t^\prime)
\label{eq:pumpedcharge}
\end{equation}

Figure~\ref{fig:berry}(c) shows the current and pumped charge \red{for pumping cycle $T=40\hbar/E_{R}$ in the Rice-Mele pumping potential (the band gap is $\sim1.5E_{R}$).} The sudden onset of pumping causes high frequency oscillations of the current. The pumped charge is quantized at times that are multiples of the cycle time $T$. Our calculations show that modifying the quantum pump by changing $V_{1}$ and $V_{2}$ results in a different current $J(t)$, however, the pumped charge remains quantized. \red{One thus realizes} a topological pump, not relying on a TB approximation, nor the details of the pumping protocol. 

\begin{figure}[t]
\centering
  \includegraphics[width=7.5cm]{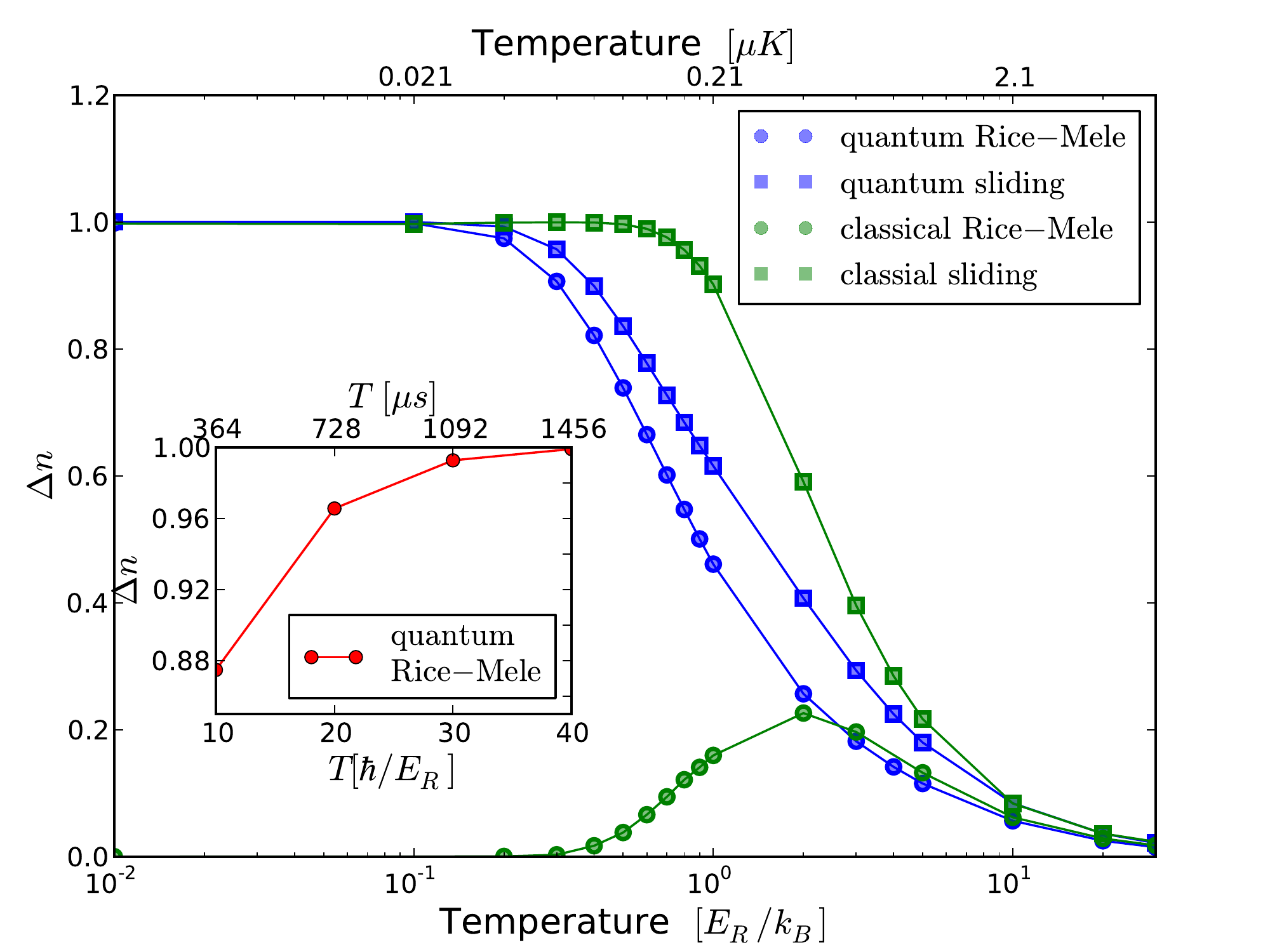}
\caption{The pumped charge after one cycle vs temperature for the quantum and classical case. Quantization of the pumped charge is visible for temperature lower than the band gap. For classical case there is no such topological protection and the pumped charge depends on the pumping protocols. Inset shows the non-adiabatic  (finite-pumping time) effect on the pumped charge of a quantum Rice-Mele pump. The upper axis shows the realistic temperature and time estimated for $^{40}$K atoms in a $d=532$ nm laser. }
\label{fig:classicalvsquantum}
\end{figure}

Figure~\ref{fig:classicalvsquantum} shows  non-adiabatic  and finite temperature effects on the  quantization of pumped charge.  Quantization is precise for slow pumping and low temperature compare to the band gap. Considering $^{40}$K atoms and $d=532$ nm, one has $\hbar/E_{R}=36.4\mu$s and $E_{R}/k_{B}=0.21\mu$K. Thus for a pumping period longer than $50\hbar/E_{R}\approx 2$ ms and initial temperature lower than $0.1 E_{R}/k_{B} \approx 20 n$K, the pumped charge is quantized to within $0.2\%$. Such a pump is feasible within current experimental abilities. 



To demonstrate the importance of quantum mechanics for  topological protection, we  examine classical pumping in the same lattice potential. For a sliding lattice ($V_{1}=0,V_{2}=4E_{R}$), both the quantum and classical pump transfer unit charge in one cycle. However, mapping the classical problem to a classical pendulum (see supplementary material) shows that this is accidental and the pumped charge is not exactly quantized. This accidental quantization is  removed by changing the atom mass, lattice constant or the pumping potential. Figure~\ref{fig:classicalvsquantum} shows that the pumped charge drops to close zero for the classical Rice-Mele pump at low temperature. It is because of the potential minima felt by classical particle does not shift in space. On the contrary, the quantization in the quantum case is protected by an energy gap and survives as one distorts the pumping potential to the Rice-Mele model. The difference between classical and quantum behavior is due to the absence of Berry phases and energy gaps in classical dynamics. This comparison  highlights the importance of quantum effects for topological protection. 

At finite temperatures,  quantization in the quantum pump remains stable for temperature smaller than the energy gap (see Fig.~\ref{fig:classicalvsquantum}). The quantum to classical transition is determined by the condition $n\lambda \ll 1$, where $\lambda=\hbar \sqrt{\frac{2\pi\beta}{m}}$ is the thermal de Broglie wavelength, $\beta$ is the inverse temperature and $n$ is average density of the system. Classical behavior dominates when $\beta^{-1} \gg \frac{4(nd)^{2}}{\pi} E_{R}$. For a shallow optical lattice $nd\sim 1$,  the quantum to classical crossover happens at temperatures much larger than $E_{R}$. For $^{40}$K atoms and $d=532$ nm  the whole temperature region of the  quantum to classical  crossover  in Fig.~\ref{fig:classicalvsquantum} can be achieved in experiments.

\begin{figure}[t!]
\centering
  \includegraphics[width=8.5cm]{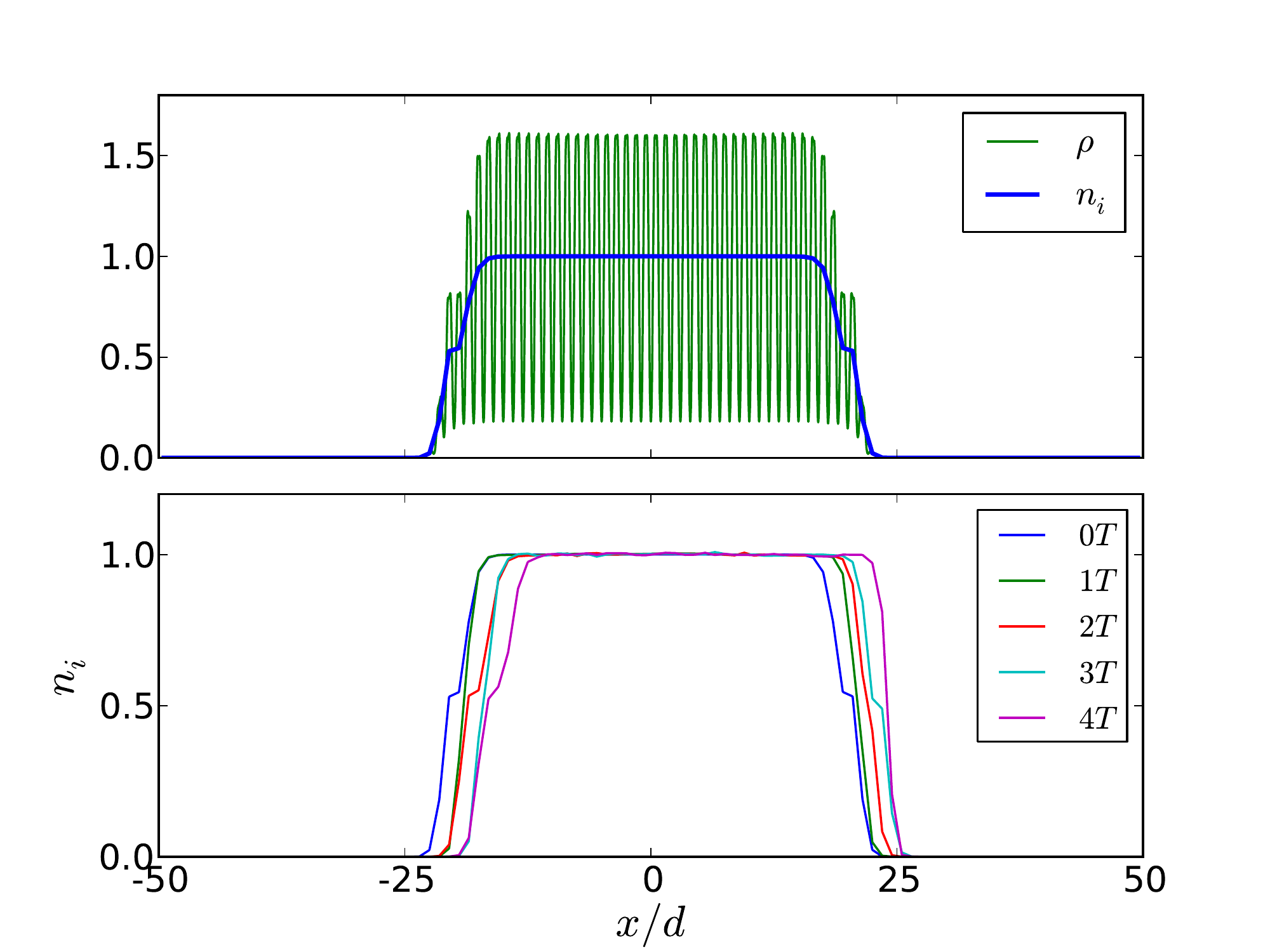}
\caption{Upper panel: initial density distribution in a harmonic trap with $\omega_\mathrm{T}=0.03E_{R}$, $V_{1}=V_{2}=4E_{R}$ and particle number $N=40$. The green curve shows the continuous space density, while the blue curve shows the  occupation number integrated over each unit cell. Lower panel: occupation number after several cycles of pumping with $T=40\hbar/E_{R}$. The cloud shifts to the right and the centre of mass position is quantized, see Fig. ~\ref{fig:com}. }
\label{fig:rho}
\end{figure}

\paragraph{Trapped system} 
To connect to \red{real experimental situations} we link the quantization to a simple physical observable: the center-of-mass of a cloud in a harmonic trap $V_\mathrm{trap}(x)= \frac{1}{2}m\omega_\mathrm{T}^{2}x^{2}$ which varies slowly compared to the optical lattice. In Fig.~\ref{fig:rho}(a) we show the initial ground state density distribution $\rho(x,t=0)$ in a trap with frequency $\omega_\mathrm{T}=0.03E_{R}$. In order to clearly see the nature of the state of trapped gas we integrate the density over each unit cell, arriving at site occupations

\begin{eqnarray}
n_{i}(t) &=& \int_{\Omega_{i}}\!\mathrm{d} x\, \rho(x,t),
\end{eqnarray}
which are show as blue lines in Fig. ~\ref{fig:rho}(a). We see a band insulator ($n_{i}=1$) in the center of the trap with very small metallic wings.

Calculating the time evolution we show in Fig. ~\ref{fig:rho}(b) the occupation number after multiple pumping cycles for $V_{1}=V_{2}=4E_{R}$, and $T=40\hbar/E_{R}$. We observe that the cloud  shifts to the right under the action of the pump.  \red{To reveal the topological nature of this drift, we show the center of mass (COM) of the cloud 
\begin{eqnarray}
\langle  x(t) \rangle & =&  \frac{1}{N}\int _{-\infty}^{\infty}\!\mathrm{d} x\,  \rho(x,t) x 
\end{eqnarray}
encodes the topological pumped charge $\Delta n$}

\begin{equation}
\langle  x \rangle/d = \Delta n,
\label{eq:com}
\end{equation}
Eq.(\ref{eq:com}) links the pumped charge $\Delta n $ with the physical observables $\langle  x \rangle$.   Experimentally, the COM position $\langle  x \rangle$ can be measured precisely, either by in situ measurement of the density distribution or \red{deduced indirectly} from time-of-flight imaging \cite{Fertig:2005p7886, Strohmaier:2007p10336}. The topological pumping effect can then be identified as a quantization of COM position at multiple pumping cycles. \red{To proof Eq.(\ref{eq:com}), we multiply $x$ to both sides of continuity equation and then integrate over spacetime, noticing that the current dies out at infinity for a trapped system.}

%

\begin{figure}[t!]
\centering
  \includegraphics[width=8cm]{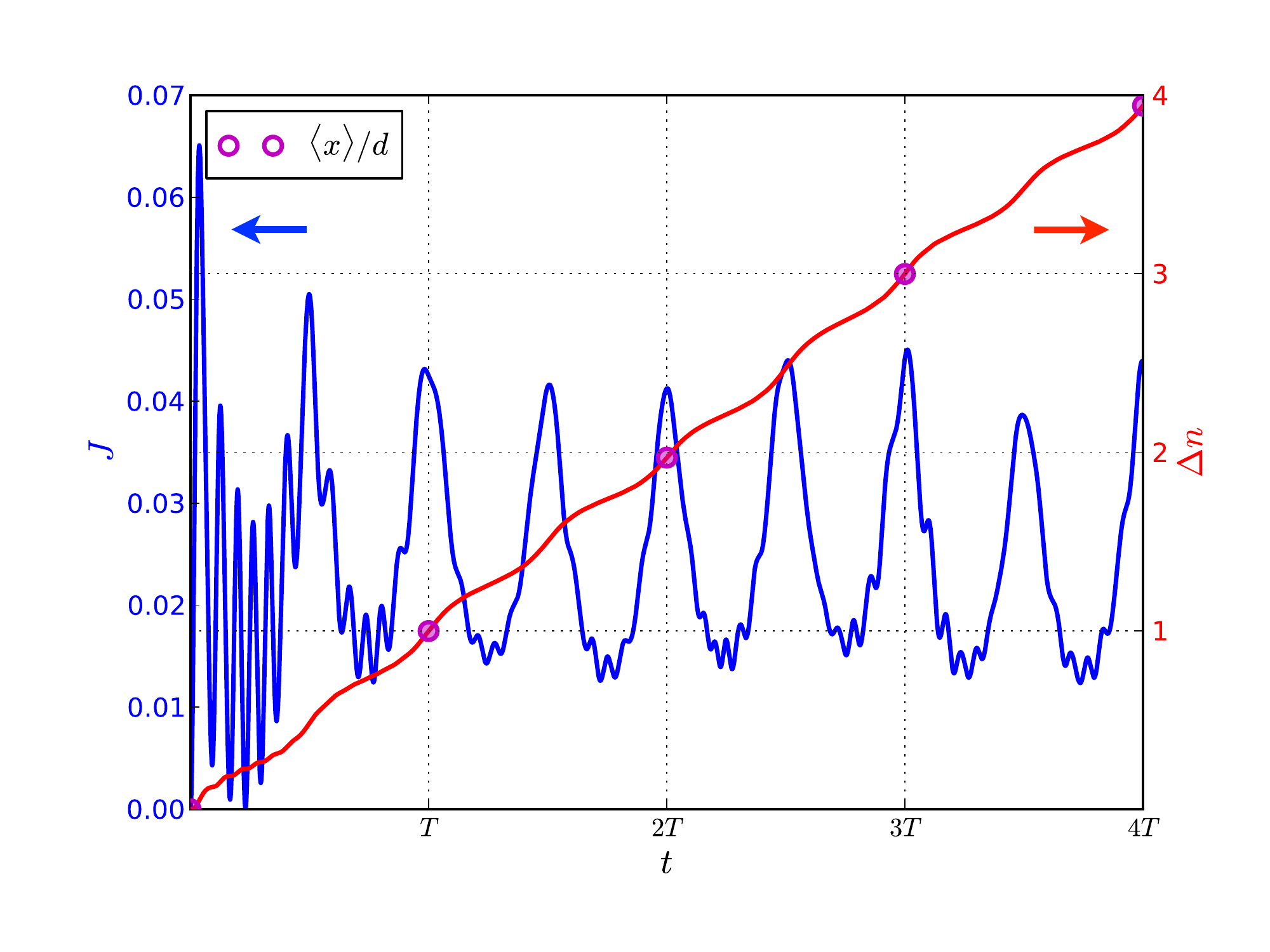}
\caption{Total current (blue line), pumped charge (red line) and center of mass (pink circles) of the atomic cloud in a harmonic trap. The COM shift equals to the pumped charge and is quantized at integer cycles. The pumping parameters are the same as Fig. ~\ref{fig:rho}}
\label{fig:com}
\end{figure}

Figure~\ref{fig:com} shows the total current $J$, pumped charge $\Delta n$ and COM $\langle x \rangle$  in a trap. The relationship Eq.(\ref{eq:com}) is evident from the plot and we clearly see quantization of the pumped charge and COM at every full pumping cycles. The similarities between Figs.~\ref{fig:com} and \ref{fig:berry} shows that external trap and finite size of the atomic cloud does not affect the precise quantization of the pumped charge. The observation of this effect in cold atom systems is thus highly feasible with in situ imaging techniques for the atomic cloud~\cite{Bakr:2010p33127, Sherson:2010p21292, Weitenberg:2011p29386}.

Finite-size effects and metallic edges will, in principle, give a non-quantized value of the pumped charge. Any such deviation from an integer value is, however, not visible in our simulations and will be even smaller in the experimental situation where the trap is larger and finite size effects are thus smaller. While topological pumping is stable against weak interactions \cite{Niu:1984p61423}, interaction effects can also easily be avoided by using spin-polarized atoms.




\paragraph{Conclusion}
We have proposed a realistic experimental setup to realize the topological pumping of cold atoms. Our setup naturally interpolates between  sliding potentials and the Rice-Mele model \cite{Rice:1982p61164} commonly studied in condensed matter physics. The quantization of the pumped charge can be observed from a quantization of the center of mass motion of the atomic cloud, is independent of details of the pumping protocol and robust with respect to nonzero temperature. The experimental observation of  topological pumping in cold atoms will be a big step towards exploring topological states and nonequilibrium dynamics in cold atom systems. 

As further steps, interactions on a fractionally occupied lattice may open up an energy gap and one could pump a fractional charge in each pumping cycle.  With two spin-species, it will be interesting to see  $Z_{2}$ spin pumping where the Wannier center of two time-reversal-symmetrical states split and exchange \cite{Fu:2006p3991, Yu:2011p37026}.


\paragraph{Acknowledgment} The work is supported by the Swiss National Science Foundation through the NCCR QSIT and the European Research Council. X.D. is supported by National Science Foundation of China. We thank Thomas Uehlinger, Jean-Philippe Brantut,  Ulrich Schneider and  Christian Gross for helpful discussions. The simulations were performed on the Brutus cluster at ETH Zurich.  

\bibliographystyle{apsrev4-1}
\bibliography{Pumping}

\clearpage
\section{Supplemental Material}

\subsection{Simulation Method and Chern number calculation}

To obtain the starting state we solve the time-independent Schr\"{o}dinger equation $H(k_{x},t=0) |\psi_{k_{x}\rangle } = E_{k_{x}} |\psi_{k_{x}}\rangle $ in a plane-wave basis of $1024$ plane waves at each of $N_{k}=128$ points in the Brillouin zone. We then fill the energy bands according to the Fermi-Dirac distribution and  evolve these wave functions in time using a second-order Trotter decomposition of the Hamiltonian into kinetic and potential terms with time step $\Delta t =0.01$. For the simulation of a trapped system, we consider a lattice with $100$ unit cells and  $N=40$ spinless fermions inside it. The number of plane wave basis are enlarged to $4000$ in this case.

From the wave-function $\psi_{k_{x}}(x,t)= \langle x| \psi_{k_{x}}\rangle$ we calculate  the local current
\begin{equation}
j(x,t)=\frac{\hbar}{2mi} \frac{1}{N_{k}}\sum_{k_{x}}(\psi_{k_{x}}^{\ast} \nabla \psi_{k_{x}}-c.c.)
\end{equation}

In the infinite system, the global current is defined as the current flowing through a unit cell can be calculates by integration over a  unit cell $\Omega$:
\begin{equation}
J(t)=\frac{1}{d}\int_{\Omega}\!\mathrm{d} x\, j(x,t) \label{eq:current}.
\end{equation}
For a trapped system, on the other hand we define the current by integrating over the whole space and normalized by the total particle number $N$:
\begin{equation}
J(t) = \frac{1}{Nd}\int_{-\infty}^{\infty} \!\mathrm{d} x\, j(x,t)
\label{eq:trapcurrent}.
\end{equation}

The {\it Chern number} of $H(k_{x},t)$ is calculated as 
\begin{equation}
C = \frac{1}{2\pi} \int_{0}^{T}\mathrm{d} t \int_{0}^{2\pi} \mathrm{d} k_{x}\, \mathcal{F}(k_{x},t) 
\end{equation}
where $\mathcal{F}(k_{x},t) = \partial_{t} A_{k_{x}} -\partial_{k_{x}}A_{t}$ is the Berry curvature and $A_{t(k_{x})} =-i \langle \psi_{k_{x}}(t) |\partial_{t(k_{x})}|\psi_{k_{x}}(t)\rangle $ is the Berry connection \cite{Xiao:2010p19709}.

\subsection{Classical charge pumping}
For classical pumping we solve the Newton's equation 
\begin{equation}
m \ddot{x} = -\frac{\partial V_\mathrm{OL}(x,t)}{\partial x}
\label{eq:newton}
\end{equation}
with initial condition that the particle resides in the minimum of the potential well with zero velocity. In the case of a sliding potential ($V_{1}=0$) it reads  
\begin{equation}
m \ddot{x} =  - \frac{\pi V_{2}}{d} \sin \left(  \frac{2\pi t}{T}-\frac{2\pi x}{d}\right)
\label{eq:slidingnewton}
\end{equation}
with  initial conditions $x(0)=0.5d$ and $\dot{x}(0)=0$. In the reference frame of the sliding lattice,  we define a new variable $\theta = 2\pi ( \frac{x}{d}-\frac{t}{T}) -\pi$ and by introducing $\omega_{0}= \sqrt{\frac{2\pi^{2} V_{2}}{m d^{2}}}$. Eq. (\ref{eq:slidingnewton}) becomes a simple pendulum equation
\begin{equation}
 \ddot{\theta} + \omega_{0}^{2} \sin \theta = 0  
\end{equation}
with the initial conditions $\theta(0) = 0$ and $\dot{\theta}(0) = -\frac{2\pi}{T}$. This equation describes a pendulum initially starting from the equilibrium position with an angular velocity inversely proportional to the pumping cycle $T$. 

The change of the particle position in the lab frame of reference is $\frac{x}{d}-\frac{1}{2}=\frac{t}{T} + \frac{\theta}{2\pi}$.  It contains a linear drift term $t/T$ which is naturally quantized at multiple cycle and an oscillatory term $\theta/2\pi$. $\theta$ oscillates with the pendulum period $\frac{4}{\omega_{0}}K(\sin\frac{\theta_\mathrm{max}}{2})$, where $K(x)$ is the complete elliptic function of the first kind and $\theta_\mathrm{max}=\arccos(1-\frac{2\pi^{2}}{\omega_{0}^{2}T^{2}})$. In general the pendulum does not return to its equilibrium position $\theta=0$ at multiples of $T$, and thus there is no quantization in the classical case. 
 
If $T<\sqrt{\frac{m}{2V_{2}}}d$, the pendulum swings around the pivot, cancelling the linear drift.  This corresponds to a fast pumping where the particle can not follow the sliding lattice.  In the opposite case $T>\sqrt{\frac{m}{2V_{2}}}d$, the angle $\theta$ oscillates between $\pm \theta_\mathrm{max} $. In the extreme case $T\rightarrow\infty$, $\theta_\mathrm{max}\rightarrow 0$ we obtain adiabatic pumping where the particle follows the sliding lattice. For slow pumping,  expand $\theta_\mathrm{max}$  as leading order $\frac{2\pi}{\omega_{0}T}$, implying that the deviation from exact quantization is inversely proportional to $T$. This is different from the quantum case where  adiabaticity is protected by the energy gap and the quantization is exponentially accurate as $T$ increases. 

\begin{figure}[h!]
\centering
  \includegraphics[width=7cm]{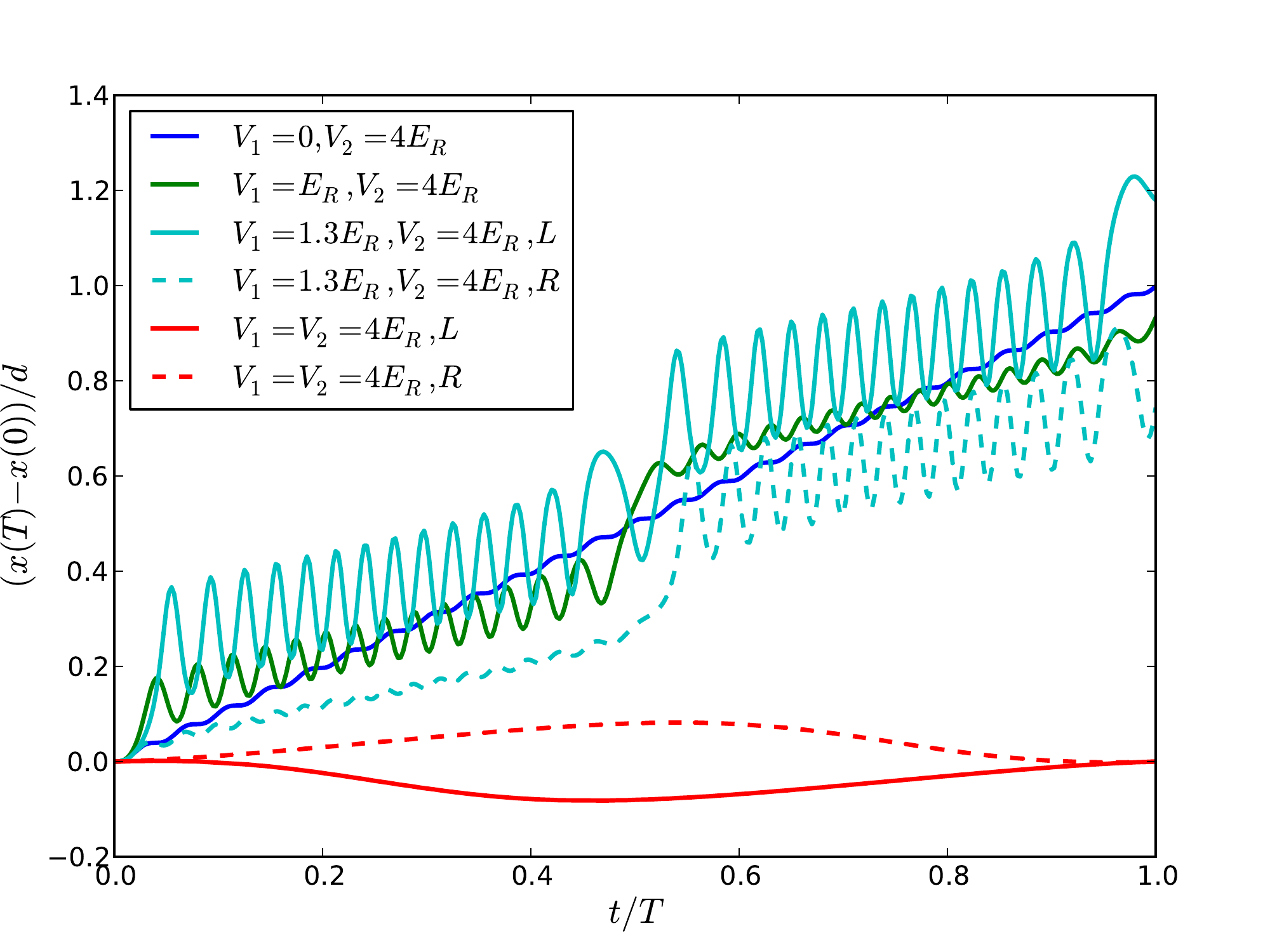}
\caption{The change of the position of a classical particle in different pumping potentials. It is in general not quantized at multiple pumping cycle. For $V_{1}>V_{2}/4$ there are two degenerate initial equilibrium positions ($x_{L},x_{R}$). They lead to different trajectories.}
\label{fig:classicalpumping}
\end{figure}

When $V_{1}\neq0$, the system acts like a periodically driven pendulum and shows rich dynamical properties. In particular, for $V_{1}>V_{2}/4$ the initial potential has two degenerate minimals $(x_{L},x_{R})$ and they lead to different trajectories of the particle. Fig. \ref{fig:classicalpumping} shows the position of particles for various pumping potentials. In general the pumped charge is not quantized at multiple cycles because there is no topological protection.

To study  finite-temperature effects in  classical  dynamics, we sample the classical trajectories according to  the Boltzmann weight 
\begin{equation}
\bar{O}= \frac{\iint e^{-\frac{E(x,v)}{k_{B}T} } O(x(t),v(t),t) \, \mathrm{d} x \mathrm{d} v }{\iint e^{-\frac{E(x,v)}{k_{B}T} } \, \mathrm{d} x \mathrm{d} v}
\end{equation}
for an arbitrary observable $O$ by integrating Eq.(\ref{eq:newton}) from initial positions $x$ and velocities $v$ and using Boltzmann weights based on the energy $E(x,v)= \frac{1}{2}mv^{2} + V_\mathrm{OL}(x, t=0)$.

\subsection{Topological equivalence between the sliding potential and the Rice-Mele pumping potential}

\begin{figure}[h!]
\centering
  \includegraphics[width=8cm]{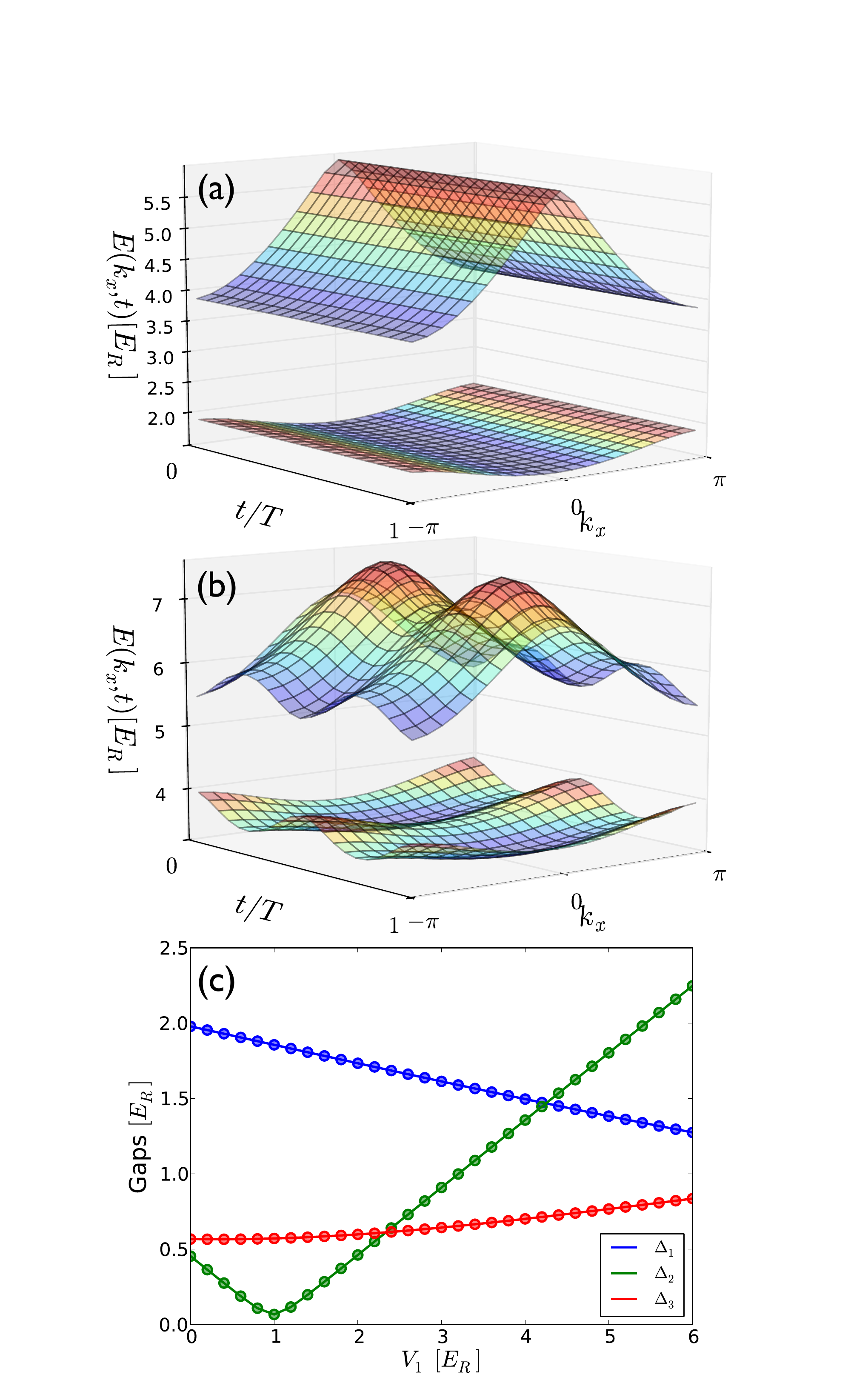}
\caption{Energy bands of (a). the sliding potential and (b). the Rice-Mele pumping potential. (c). Band gaps versus $V_{1}$ with $V_{2}=4E_{R}$ fixed. The gap between the lowest and the second band ($\Delta_{1}$) does not close. While the gap between the second and the third band ($\Delta_{2}$) closes and reopens at $V_{1}=1E_{R}$. This indicates a topological phase transition and changes the Chern number of the second band.  }
\label{fig:band}
\end{figure}

Figure \ref{fig:band}(a) and (b) shows the lowest two energy bands of the sliding potential  ($V_{1}=0$,$V_{2}=4E_{R}$) and Rice-Mele pump ($V_{1}=V_{2}=4E_{R}$). Both of them are gapped and can be adiabatically connected by increasing $V_{1}$ without closing the band gap $\Delta_{1}$ as shown in Fig. ~\ref{fig:band}(c). The two pumping protocols are thus equivalent for the lowest band. However, the band gap $\Delta_{2}$ between the second and the third band  closes when increasing $V_{1}$, indicating that the equivalence does not hold for higher bands.

\begin{figure} [tbh]
\centering
  \includegraphics[width=7.5cm]{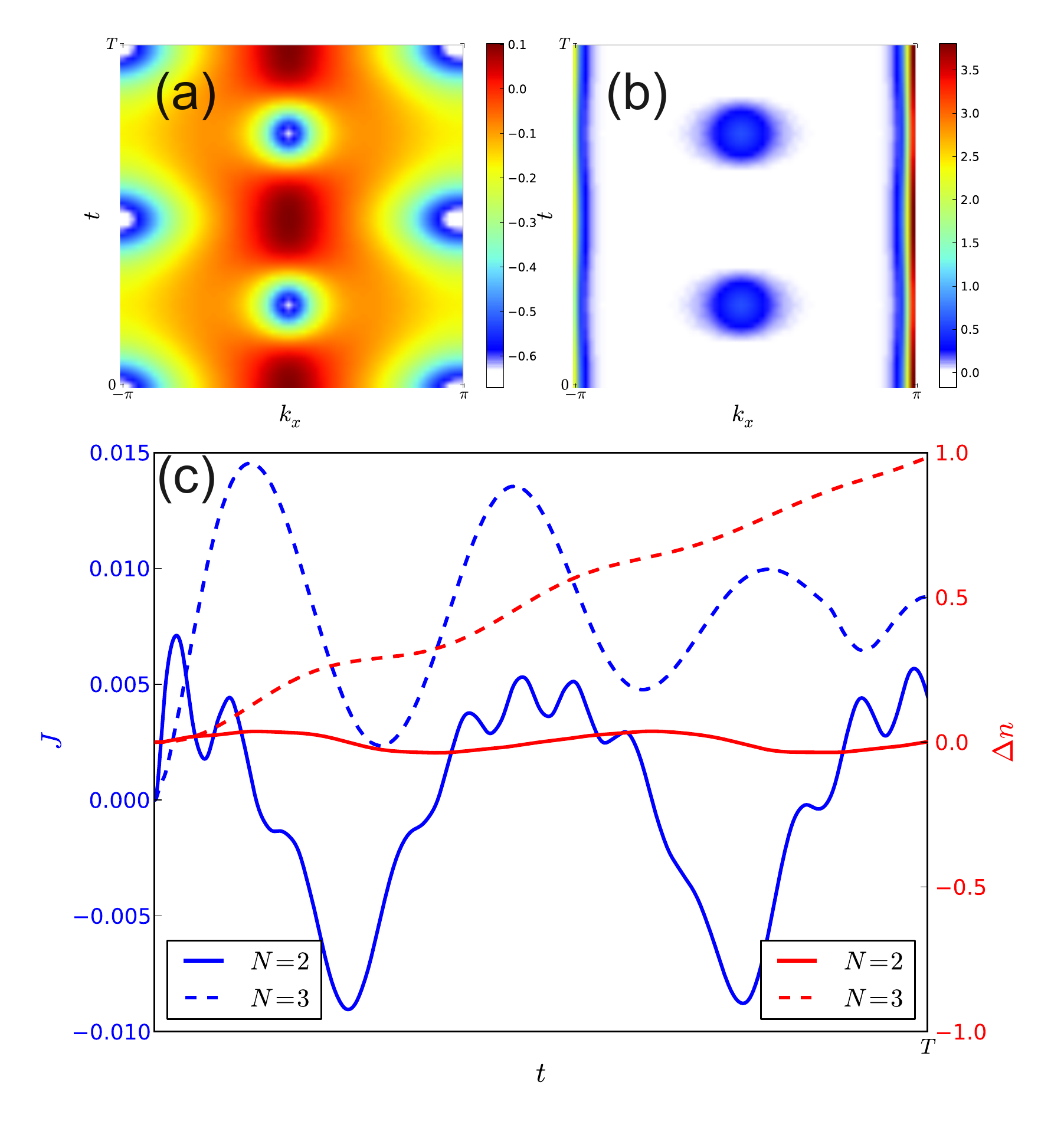}
\caption{(a-b). Berry curvatures of the second and third band of the Rice-Mele lattice ($V_{1}=V_{2}=4E_{R}$). Their Chern numbers are $-1$ and $1$ respectively. (c). Total current (blue) and pumped charge (red) for filled lowest two ($N=2$, solid lines) and three ($N=3$, dashed lines) energy bands, $T=40\hbar/E_{R}$.}
\label{fig:higherband}
\end{figure}

\subsection{Band insulator with higher fillings}
Figure \ref{fig:higherband} shows the Berry curvatures for higher bands of the Rice-Mele potential. Their corresponding Chern numbers are $-1$ and $1$ respectively. Thus for a band insulators with 2 and 3 atoms per unit cell, the pumped charge in one cycle is 0 and 1 respectively. This is in contrast to the sliding potential, where the pumped charges are 2 and 3, because each band contributes Chern number 1. This result agrees with  considerations based on the Floquet operator \cite{Kitagawa:2010p55069}. Experimentally, the adiabatic condition is more restricted for higher bands, because the band gaps are smaller. 

%


\subsection{Proof of the Eq. (\ref{eq:com})}
We multiply $x$ to both sides of the continuity equation $\frac{\partial \rho}{\partial t} = - \nabla j$ and integrate over space, 

\begin{eqnarray}
\int_{-\infty}^{\infty} \!\mathrm{d} x\, \frac{\partial \rho}{\partial t}  x = -\int_{-\infty}^{\infty} \!\mathrm{d} x\,\nabla j  x   =  \int_{-\infty}^{\infty} \!\mathrm{d}x \,j
\label{eq:proof}  
\end{eqnarray} 
For the  integration by parts we have used the fact that current dies out at infinity. Integrate both sides of Eq. (\ref{eq:proof}) from $0$ to $t$,  

%

\begin{eqnarray}
 \int_{-\infty}^{\infty} \!\mathrm{d} x\, \left[\rho(x,t)-\rho(x,0)\right]  x  =  \int_{0}^{t} \!\mathrm{d} t^{\prime}\,\int_{-\infty}^{\infty} \!\mathrm{d} x\, j(x,t^{\prime})  
\end{eqnarray} 
Use the fact that the initial COM is zero and Eq.(\ref{eq:pumpedcharge}), one proofs Eq. (\ref{eq:com})

\subsection{Effect of trap on the quantization}

\begin{figure}
\centering
  \includegraphics[width=7.5cm]{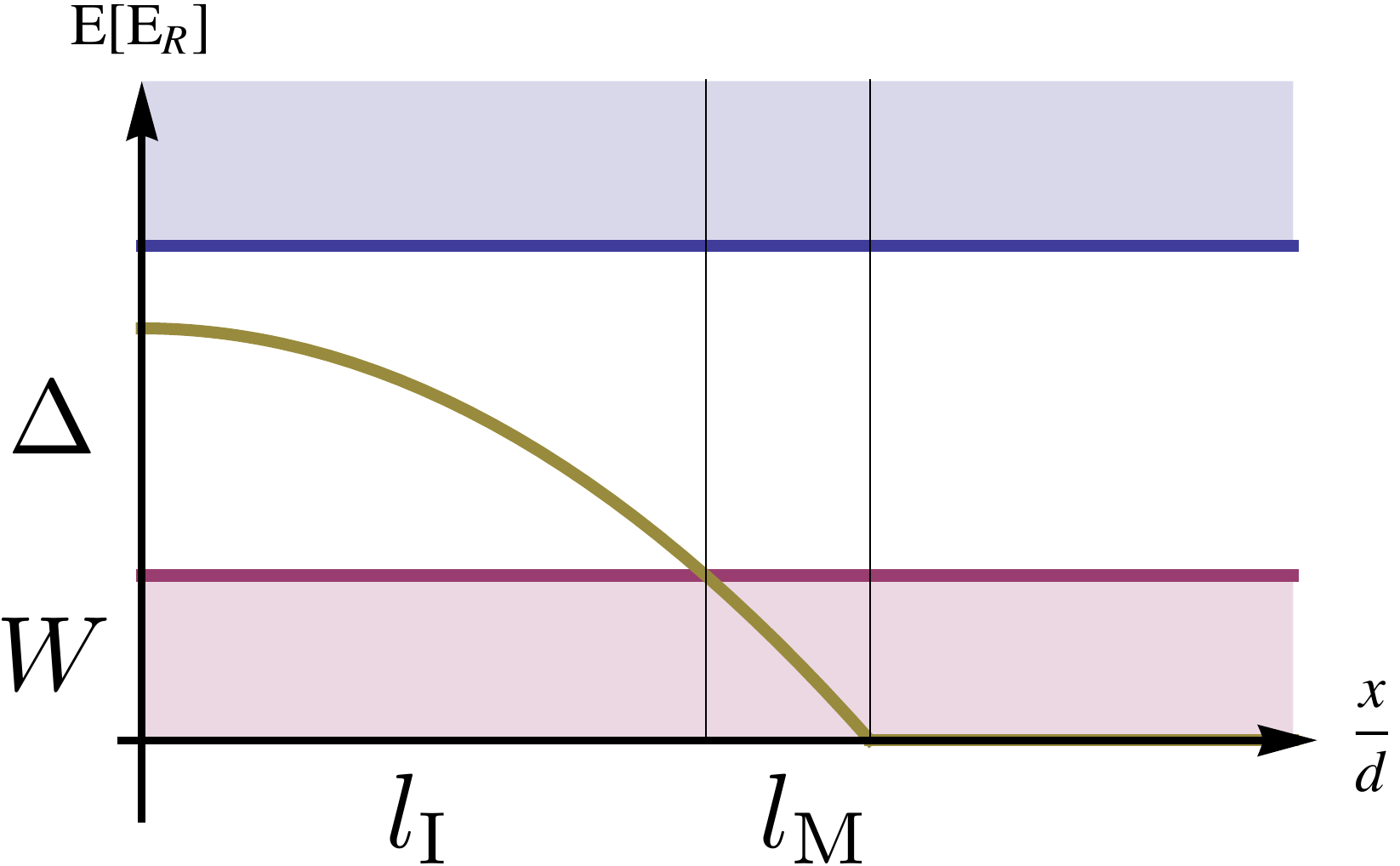}
\caption{LDA estimation for the length of metallic wings $l_{M}$ versus the band insulators $l_{I}$. $W$ denotes the band width of the lowest band, $\Delta$ denotes the band gap to the higher band. Golden line indicates the local chemical potential in the trap.}
\label{fig:lda}
\end{figure}

Under the local-density-approximation (LDA), the chemical potential in the trap is $\mu(x) = \mu_0 - \frac{1}{2}m\omega_\mathrm{T}^{2}x^{2}$. We want to estimate the length of the band insulator plateau ($l_I$) and metallic wings ($l_M$). Their ratio affects the quantization of center-of-mass. Denotes $W$ and $\Delta$ as the band width of the lowest band and the band gap to the higher band. Assume the chemical potential in the center of the trap reaches $W+\Delta$, one has (shown in Fig.~\ref{fig:lda})

\begin{eqnarray}
W + \Delta = \mu(0)  \\
W = \mu(l_I)\\
0 = \mu(l_I + l_M)
\end{eqnarray}

One has $l_M/l_I = \sqrt{W+\Delta}/\sqrt{\Delta} -1$. The ratio is proportional to $ \sqrt{W/\Delta}$ when $W/\Delta \ll 1$. Notice that it does not depend on the strength of the trap. For the data showed in the paper (Fig.~\ref{fig:rho} and Fig.~\ref{fig:com}), we have $W/\Delta \sim 1/3$, and one already has a precision of quantization about $0.2\%$. To improve the accuracy of quantization, one just needs to increase the optical lattice depth and will have a smaller $W/\Delta$ ratio.  

\end{document}